\documentclass[letterpaper,english]{emulateapj}
\usepackage[T1]{fontenc}
\usepackage[latin9]{inputenc}
\usepackage{units}
\usepackage{amssymb}
\usepackage{graphicx}

\makeatletter

\shorttitle{X-Raying an Accretion Disk in Realtime during a Superburst}
\shortauthors{Keek, Ballantyne, Kuulkers, \& Strohmayer}

\makeatother

\usepackage{babel}
\begin{document}

\title{X-Raying an Accretion Disk in Realtime:\\
 the Evolution of Ionized Reflection during a Superburst from 4U\,1636--536}

\author{L.~Keek and D.\,R.~Ballantyne}

\affil{Center for Relativistic Astrophysics, School of Physics, Georgia
Institute of Technology, 837 State Street, Atlanta, GA 30332-0430,
USA}

\email{l.keek@gatech.edu}

\author{E.~Kuulkers}

\affil{European Space Astronomy Centre (ESA/ESAC), Science Operations Department,
28691 Villanueva de la Cañada, Madrid, Spain}

\author{\and T.\,E.~Strohmayer}

\affil{X-ray Astrophysics Lab, Astrophysics Science Division, NASA's Goddard
Space Flight Center, Greenbelt, MD 20771, USA}
\begin{abstract}
When a thermonuclear X-ray burst ignites on an accreting neutron
star, the accretion disk undergoes sudden strong X-ray illumination,
which can drive a range of processes in the disk. Observations of
superbursts, with durations of several hours, provide the best opportunity
to study these processes and to probe accretion physics. Using detailed
models of X-ray reflection, we perform time resolved spectroscopy
of the superburst observed from 4U~1636--536 in 2001 with \emph{RXTE}.
The spectra are consistent with a blackbody reflecting off a photoionized
accretion disk, with the ionization state dropping with time. The
evolution of the reflection fraction indicates that the initial reflection
occurs from a part of the disk at larger radius, subsequently transitioning
to reflection from an inner region of the disk. Even though this superburst
did not reach the Eddington limit, we find that a strong local absorber
develops during the superburst. Including this event, only two superbursts
have been observed by an instrument with sufficient collecting area
to allow for this analysis. It highlights the exciting opportunity
for future X-ray observatories to investigate the processes in accretion
disks when illuminated by superbursts.
\end{abstract}


\section{Introduction}

Accretion onto neutron stars in low-mass X-ray binaries provides
the fuel for runaway thermonuclear burning that powers Type I X-ray
bursts \citep[e.g.,][]{Lewin1993}. Aside from many brief ($10-100\,\mathrm{s}$)
bursts, a rare class of X-ray bursts is observed that last many hours:
superbursts \citep{Cornelisse2000,Strohmayer2006}.  They originate
from the same systems as the short bursts, but are thought to be due
to a carbon flash $\sim100\,\mathrm{m}$ below the neutron star surface
\citep{Cumming2001,Strohmayer2002}. Although superbursts can reach
peak brightnesses similar to those of short bursts, their long durations
provide for spectra with much higher signal-to-noise. Only two instances,
however, were detected with an instrument capable of collecting such
spectra: the Proportional Counter Array \citep[PCA;][]{Jahoda2006}
on the \emph{Rossi X-ray Timing Explorer} \citep[RXTE;][]{Bradt1993}
observed a superburst from 4U\,1820--30 in 1999 \citep{Strohmayer2002}
and from \object[4U 1636-536]{4U\,1636--536} in 2001 \citep{Strohmayer2002a,2004Kuulkers}.

\citet{Keek2014sb1} performed a detailed time-resolved spectral analysis
of the 4U\,1636--536 superburst using a phenomenological spectral
model. This revealed that the persistent flux increased in brightness
and shifted to lower energies during the superburst, returning to
pre-superburst values in the tail. Furthermore, the spectra exhibit
an emission line and absorption edge close to $6.4\,\mathrm{keV}$.
Similar features in the spectra of the superburst from 4U\,1820--30
\citep{Strohmayer2002} were found to be consistent with reflection
of a blackbody spectrum off a photoionized accretion disk \citep{Ballantyne2004}.
The evolution of the reflection features suggested the temporary disturbance
of the inner disk during the burst. Several physical scenarios were
put forward to describe the behavior of the superburst--disk interaction,
including Poynting-Robertson drag, radiatively or thermally driven
winds, and an evolving disk geometry \citep{Ballantyne2005}. All
processes are expected to play a role, but their relative importance
could not be established.

In this Letter we investigate whether the spectra of the 2001 4U\,1636--536
superburst are consistent with reflection. The evolution of the ionization
state of the reflecting material is quantified, and we discuss possible
interpretations.

\section{Observations and Spectral Models}

\label{sec:Observations-and-Spectral}

The spectra, response matrices, and instrumental backgrounds created
by \citet{Keek2014sb1} are reused for this analysis. The spectra
are extracted from Standard 2 PCA data in the $3$ to $20\,\mathrm{keV}$
energy range. The superburst spans four \emph{RXTE} orbits. During
the first and second orbit, spectra were extracted in $64\,\mathrm{s}$
time intervals. In orbit $3$ the line and edge were found to be exceedingly
weak \citep{Keek2014sb1}. To maximize the reflection signal, we extract
one spectrum for orbit $3$, with an exposure time of $28$ minutes.
At this point in the superburst tail, the evolution of the spectral
parameters is sufficiently slow to warrant this approach. We exclude
orbit $4$ from our analysis, because no reflection signal is detected
at that time, as the burst illumination
has dropped substantially at the end of the superburst.

\texttt{XSPEC} version 12.8.1 \citep{Arnaud1996} is employed to analyze
the spectra. The main components of our spectral model are a power
law with a high energy cutoff (\texttt{cutoffpl} in \texttt{XSPEC})
for the persistent flux and a blackbody (\texttt{bbodyrad}) for the
neutron-star emission. Additionally, models of reflection of a blackbody
off a photoionized accretion disk are included from \citet{Ballantyne2004models},
and the grid of models is extended to include blackbody temperatures
down to $kT=0.5\,\mathrm{keV}$. They self-consistently include the
$\mathrm{Fe\, K\alpha}$ line and edge. Apart from the regular blackbody
parameters, the reflection models are a function of the ionization
parameter $\xi$: 
\begin{equation}
\xi=\frac{4\pi F}{n_{\mathrm{H}}},\label{eq:xi}
\end{equation}
with $F$ the illuminating flux at the disk and $n_{\mathrm{H}}$
the hydrogen number density. In this Letter we use the logarithm of
$\xi$, with $\xi$ expressed in units of $\mathrm{erg\, s^{-1}\, cm}$.

To include relativistic smoothing effects, such as Doppler broadening
and gravitational redshifts, the reflection components are convolved
with the \texttt{rdblur} model \citep{Fabian1989}, under the assumption
that the burst illumination pattern drops quadratically with radius.
The \texttt{cflux} model is used to determine the $0.001-100\,\mathrm{keV}$
flux of the reflected blackbody, $F_{\mathrm{refl}}$. Furthermore,
photoelectric absorption is taken into account using the \texttt{vphabs}
model with the composition of the absorbing material set to the mean
values measured by \citet{Pandel2008}  \citep[see also][]{Keek2014sb1}.
The reported uncertainties in all fit parameters are at $1\sigma$.

\section{Results}

\subsection{Superburst Spectral Fits}

\label{sub:Superburst-spectra} 

\begin{figure*}
\begin{centering}
\includegraphics{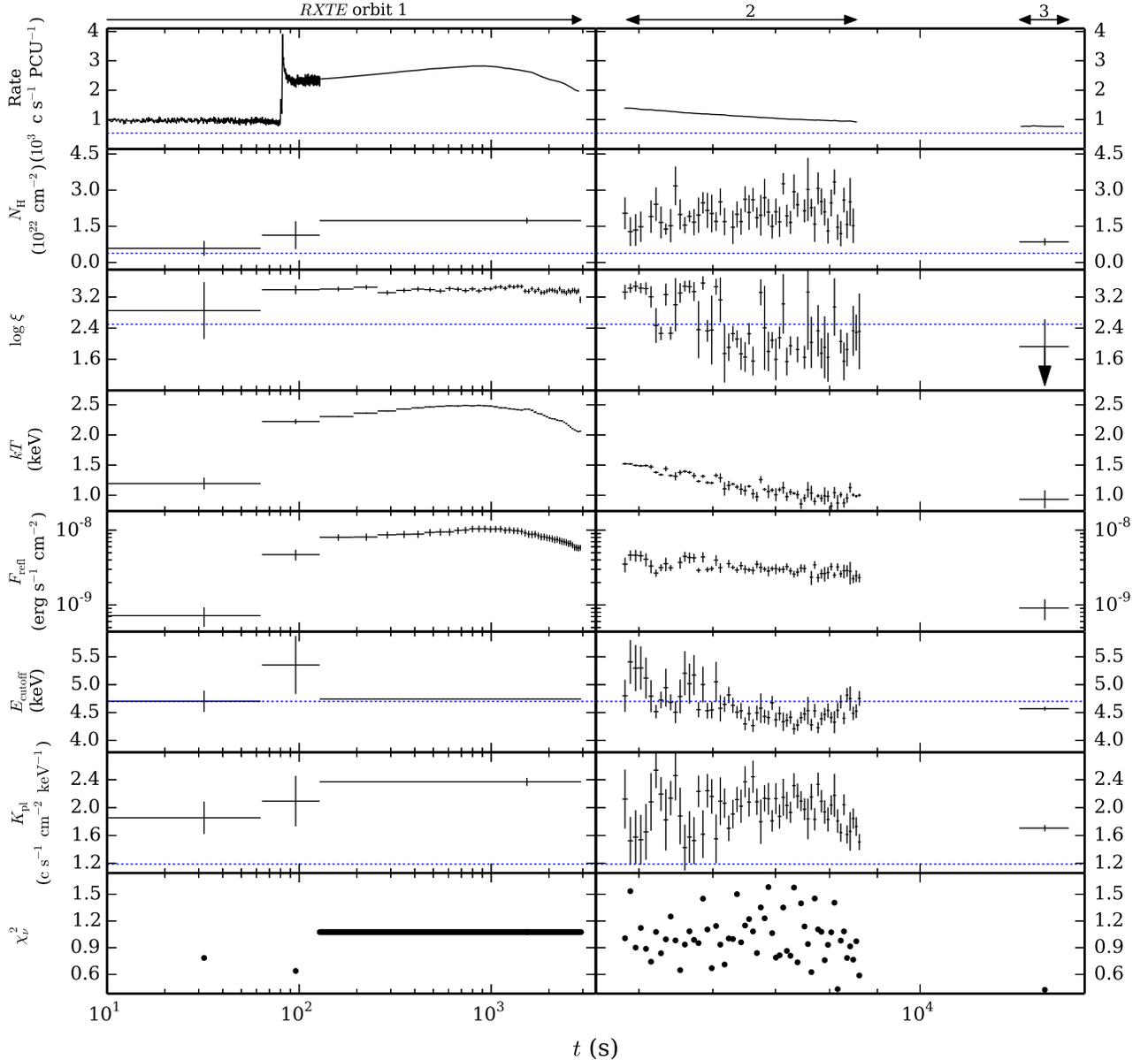}
\par\end{centering}

\protect\caption{\label{fig:fit_both}Results from the time-resolved spectral fits
as a function of the time, $t$.
At the top the count rate is shown (the initial $128\,\mathrm{s}$
at higher time resolution) as well as the numbering of the data segments
corresponding to the 3 \emph{RXTE} orbits. The $\chi_{\nu}^{2}$ value
of the combined fit is represented by a horizontal line. The superburst
is divided in a left and right panel, which have a different horizontal
scale. For each data point we indicate the width of the time interval
(horizontally) as well as the $1\sigma$ uncertainty (vertically).
A few data points are omitted where the relative error exceeds $50\%$.
Dotted lines indicate the parameter values measured in the \emph{RXTE}
orbit preceding the superburst. }
\end{figure*}

The superburst spectra are fit with the model described above (Section~\ref{sec:Observations-and-Spectral}).
We do not include a component for reflection of the power-law (Section~\ref{sub:Pre--and-post-superburst}),
because it is expected to be too weak to be detected during the burst
\citep[see][]{Keek2014sb1}.  The parameters of the \texttt{rdblur}
smoothing model are poorly constrained by the fits.  We, therefore,
choose to fix its parameters: the disk inclination angle is fixed
to $60^{\circ}$, and its inner radius to $20r_{\mathrm{g}}$, with
$r_{\mathrm{g}}$ the gravitational radius \citep[cf.][]{Pandel2008}.
The absorption column, $N_{\mathrm{H}}$, is not fixed, but is determined
by the fit.

\citet{Keek2014sb1} found that the data allow for only a limited
number of spectral components to be constrained. This is especially
problematic now that a reflection component is added. To limit the
number of free parameters, the cutoff power-law photon index is fixed
to the pre-superburst best-fit value: $\Gamma=1.06$ \citep{Keek2014sb1}.
Furthermore, the normalizations of the spectral components cannot
be constrained independently for each time bin. A combined fit is
performed of all spectra in the first orbit, excluding the earliest
two, which cover the fast part of the rise. During this fit, the values
of $N_{\mathrm{H}}$, blackbody normalization $K_{\mathrm{bb}}$,
and cutoff power-law normalization $K\mathrm{_{pl}}$ are each set
to be the same across all spectra. Previously, these parameters were
found to typically exhibit variations of at most $2\sigma$ \citep{Keek2014sb1}.
Also, the power-law cutoff energy is fixed to the pre-superburst value:
$E_{\mathrm{cutoff}}=4.8\,\mathrm{keV}$ \citep[see][]{Keek2014sb1}.
All other free parameters are allowed to be different for each spectrum
(see Figure~\ref{fig:fit_both}). This combined fit yields $\chi_{\nu}^{2}=1.08$.
For the rest of the superburst, $\Gamma$ is again fixed, but $E_{\mathrm{cutoff}}$
is a free parameter. Moreover, $K_{\mathrm{bb}}$ is fixed to the
best-fit value from the combined fit in the first orbit: $K_{\mathrm{bb}}=38.7$
\citep[see discussion by][]{Keek2014sb1}. The $\chi^{2}$ values
obtained with this model match a distribution for acceptable fits. 

The results of both parts of the superburst fit are presented in Figure~\ref{fig:fit_both}.
After a high ionization parameter in the first orbit with a weighted
mean of $\log\xi=3.384\pm0.009$ (excluding the first $128\,\mathrm{s}$),
the fit results exhibit a transition to a lower ionization state with
a weighted mean of $\log\xi=2.03\pm0.08$ in the second half of the
second orbit%
\footnote{The models exhibit a minimum in the equivalent width of the Fe~K$\alpha$
iron line near $\log\xi=2.1$ due to Auger destruction \citep{Ballantyne2004models},
but it may not be physical \citep{Garcia2013}. When we artificially
increase the line's equivalent width to remove the minimum, the values
of the fitted parameters do not change significantly. The fits are
not very sensitive to the line strength, but more to the edge properties
\citep[see also][]{Keek2014sb1}.%
}. Rather than following the flux, $\xi$ transitions between two values.
For spectra close to the time of the transition, there are two local
minima in $\chi^{2}$ present around the higher and the lower value
of $\log\xi$, and the fits jump between these. We investigated a
wide range of variations of our spectral model and different choices
of fixed parameters, but there is no smooth transition between the
two ionization states. The time of the transition is, however, sensitive
to the amount of smoothing of the reflection component. With stronger
or with no smoothing, it occurs $\sim2\times10^{3}\,\mathrm{s}$ earlier
or later. Finally, in orbit $3$, $\xi$ cannot be fully constrained,
and we report an upper limit. 

\begin{figure*}
\begin{centering}
\includegraphics{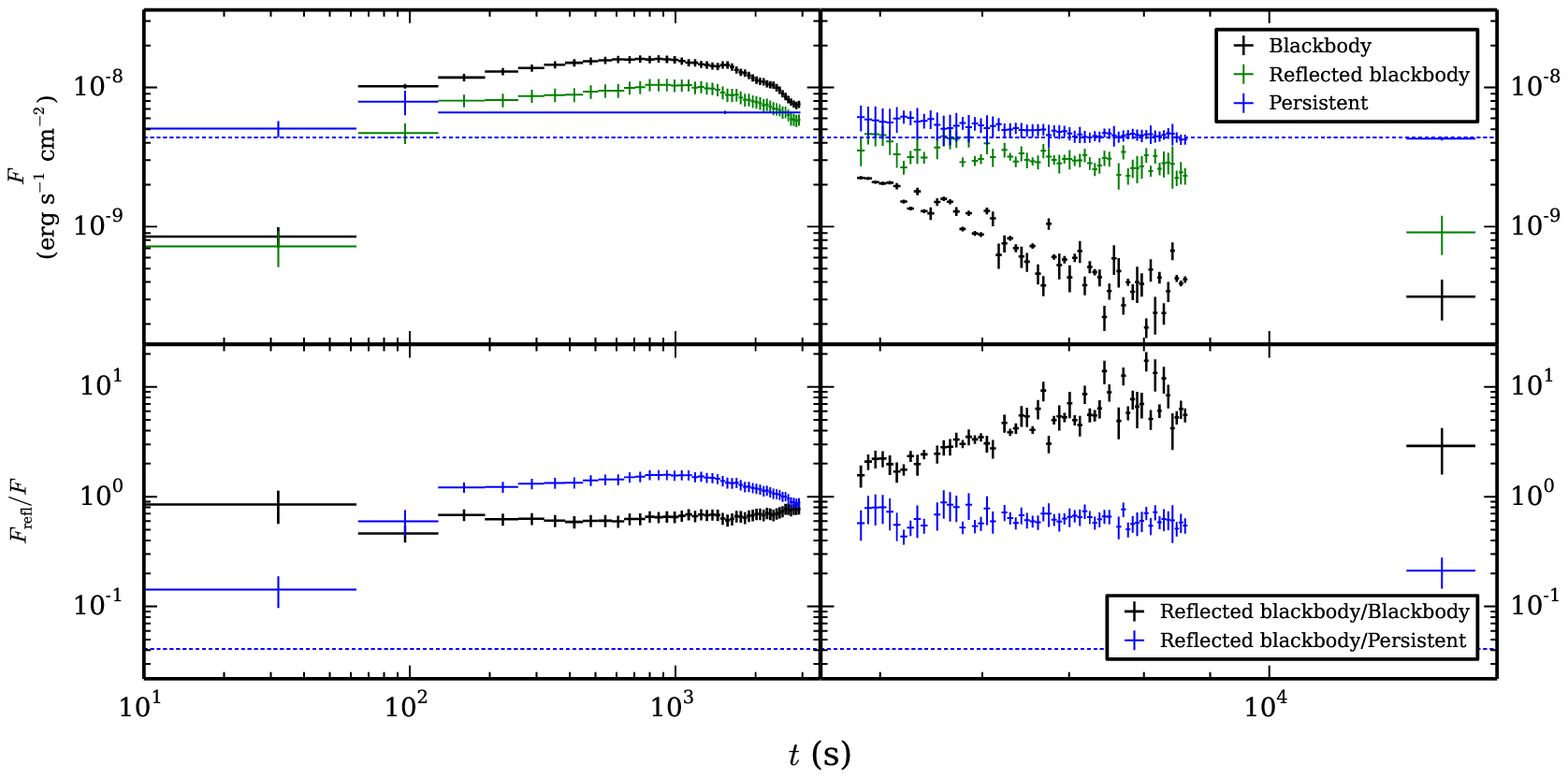}
\par\end{centering}

\protect\caption{\label{fig:fluxes}\textbf{Top}: Unabsorbed bolometric fluxes of the
blackbody and the reflected blackbody, as well as the $3-20\,\mathrm{keV}$
persistent (cutoff power law) flux. On the left the blackbody is brightest,
persistent flux weakest, and the reflection component is in between,
whereas on the right the order is reversed. The dotted line indicates
the pre-superburst persistent flux. \textbf{Bottom}: Ratio of the
flux of the reflected blackbody with respect to the blackbody and
the persistent flux. The dotted line indicates the pre-superburst
value of the ratio of the flux of a reflected power-law and the persistent
flux.}
\end{figure*}
 Figure~\ref{fig:fluxes} presents the unabsorbed flux of the different
components and the ratio of the reflected blackbody flux with respect
to the illuminating blackbody, which is a measure of the reflection
fraction. We also include the ratio of the reflected blackbody and
the persistent flux. The persistent flux is shown in the $3-20\,\mathrm{keV}$
band, whereas the (reflected) blackbody flux is bolometric, but falls
largely within the same band. Restricting the persistent flux to the
band-pass gives a better representation of which component dominates
the PCA spectra. Extrapolation to lower energies of the cutoff power
law yields a bolometric correction factor of $2.9\pm0.3$.

Whereas in the first orbit the in-band flux is dominated by the blackbody,
in the second orbit the persistent flux is stronger. In the first
orbit, the weighted mean of the ratio of the reflected and illuminating
blackbody flux is $F_{\mathrm{refl}}/F_{\mathrm{bb}}=0.665\pm0.012$.
This ratio increases in the second orbit to $F_{\mathrm{refl}}/F_{\mathrm{bb}}=6.1\pm0.3$
(weighted mean in the last $10^{3}\,\mathrm{s}$ of the orbit). Furthermore,
the reflected blackbody flux is a constant fraction of the in-band
persistent flux in the second orbit of $F_{\mathrm{refl}}/F_{\mathrm{pers}}=0.601\pm0.014$.
If reflection were dominated by the persistent flux, $F_{\mathrm{refl}}/F_{\mathrm{pers}}$
would be over an order of magnitude larger than before the superburst
(Section~\ref{sub:Pre--and-post-superburst}): much larger than the increase in 
$F_{\mathrm{pers}}$ during the superburst. We attempt to fit
the spectra in that orbit while including a reflected cutoff power
law (see also Section \ref{sub:Pre--and-post-superburst}) instead
of a reflected blackbody. The rest of the spectral model is the same,
apart from an increase of the (fixed) blackbody normalization to account
for the reflection fraction as measured in the first orbit. This does
not provide a good description of the data, especially at later times
where the blackbody flux is lowest ($\chi_{\nu}^{2}\gtrsim6$). Even
though the cutoff power law is the brightest spectral component, the
blackbody remains the dominant source of illumination for reflection.

None of the spectral parameters has an apparent correlation with the
phase of the $3.8\,\mathrm{hr}$ binary orbit \citep{Pedersen1981,Casares2006}.

\subsection{Pre- and Post-Superburst Persistent Spectra}

\label{sub:Pre--and-post-superburst}

To characterize reflection outside of the superburst, we fit the persistent
spectra in the \emph{RXTE} orbits immediately before and after the
superburst with an absorbed cutoff power-law and reflected cutoff
power-law (see \citealt{Ballantyne2012models}; grid extended for
compatibility with the \texttt{cutoffpl} model). The two components
share the photon index and cutoff energy. Smoothing by relativistic
effects is applied to the reflection component as described in Section~\ref{sub:Superburst-spectra}.
The best fit parameters of the cutoff power-law component are within
$1.7\sigma$ of those in the phenomenological fit \citep{Keek2014sb1},
with the exception of the normalization, $K_{\mathrm{pl}}$, due to
the addition of the reflection component. The in-band flux ratio of
the reflected and the illuminating cutoff power law is $F_{\mathrm{refl}}/F_{\mathrm{pl}}=(4.1\pm0.7)\times10^{-2}$
before and $F_{\mathrm{refl}}/F_{\mathrm{pl}}=(3.3\pm0.6)\times10^{-2}$
after the superburst. The material is mildly ionized with best fit
value $\log\xi=2.47_{-1.0\mathrm{p}}^{+0.12}$ before and $\log\xi=2.5_{-1.0\mathrm{p}}^{+0.2}$
after the superburst. $\xi$ is not well constrained towards lower
values, and $-1.0\mathrm{p}$ points to the pegged values at the boundary
of the domain. These values are consistent with those in the superburst
tail (Section~\ref{sub:Superburst-spectra}), although the latter
was found to be dominated by reflected blackbody emission, and the
uncertainties are rather large.

\section{Discussion}

Our analysis of the 2001 superburst from 4U~1636--536 reveals that
the spectral features identified by \citet{Keek2014sb1} are consistent
with reflection of blackbody emission from the neutron star surface
off the photoionized accretion disk. Although no other interpretations
of the data were tested, we regard burst reflection as the most likely
candidate, as it matches the strength of the emission line and absorption
edge. For example, reflection of persistent emission is much weaker
(Section \ref{sub:Pre--and-post-superburst}; see also the discussion
in \citealt{Keek2014sb1}). Here we discuss the implications of the
burst reflection interpretation.

\subsection{Transition of Ionization State and Reflection Fraction}

\label{sub:Transition-of-Ionization}

The ionization parameter, $\xi$, decreases between the peak (the
first orbit) and the tail (second half of the second orbit) of the
superburst: $\Delta\log\xi=1.35\pm0.09$. The blackbody fluxes (Figure~\ref{fig:fluxes})
yield $\log\frac{F_{\mathrm{peak}}}{F_{\mathrm{tail}}}\simeq0.9$:
a large part of $\Delta\xi$ can be explained by the decrease in the
illuminating flux alone (Equation~\ref{eq:xi}), whereas changes
in the area and density of the reflector may account for the remainder.
It does not, however, explain the non-smooth transition in $\xi$
(Section~\ref{sub:Superburst-spectra}), which suggests that two
reflection signals are present. The transition may signify a change
in which component is brightest.

The decrease in $\xi$ is accompanied by a smooth ten-fold increase
of the reflection fraction, which we define as the flux ratio of the
reflected and the illuminating blackbody. When the reflection fraction
increases, the reflected blackbody flux follows the same trend as
the persistent flux (Figure~\ref{fig:fluxes}). Both quantities may
be driven by changes in the disk. Assuming an inclination angle of
$60^{\circ}$ for 4U~1636--536 \citep[e.g.,][]{Pandel2008}, the
predicted reflection fraction for a thin disk is $\nicefrac{1}{3}$
\citep[following][]{fujimoto88apj}, which may be increased by several
tens of percent due to light bending \citep{lapidus85mnras}. The
large observed reflection fractions of up to $\sim6$, require a thick
disk that blocks part of the neutron star from the observer's view
\citep[e.g.,][]{Blackman1999}. The inner region of a thick disk may
receive the bulk of the irradiation, and dominate the late-time reflection.
At early times the superburst must, therefore, have induced a change
in the inner disk. Perhaps the disk was flattened by radiation pressure
\citep[e.g., ][]{lapidus85mnras} or thinned by a strong wind \citep[e.g.,][]{Ballantyne2005}
around the superburst peak, and the geometry reverted to a thick disk
when the burst flux dropped. Alternatively, X-ray heating could have
decreased the density of the inner disk such that it is optically
thin and does not reflect, or a part of the inner disk may be evacuated
by winds or Poynting-Robertson drag \citep[e.g.,][]{Ballantyne2005}.
Furthermore, for large values of $\xi$ reflection features disappear
\citep[e.g.,][]{Ballantyne2004models}, and part of the reflection
flux may be indistinguishable from the illuminating blackbody in the
PCA band, leading to an underestimation of the reflection fraction. 

Any of these processes could shift the location that dominates the
reflection spectrum to a region with a lower $\xi$. Detailed numerical
modeling is required to investigate the relative importance of each.
In any case, at least two reflecting regions are required to explain
the observed behavior, each with a different $\xi$ and reflection
fraction.

\subsection{Strongly Enhanced Local Absorption}

The column density of photo-electric absorption, $N_{\mathrm{H}}$,
is on average $5$ times larger during the superburst than the value
outside of bursts \citep{Pandel2008}. This suggests that in addition
to interstellar absorption, a strong local absorber developed during
the superburst. Alternative models of local absorption describe the
data equally well, including a photoionized absorption model created
using XSTAR \citep{Kallman2001xstar} and a model with partial covering.
As most absorption must take place below the PCA's band-pass, the
data do not discriminate between different models. Although the precise
$N_{\mathrm{H}}$ values are model dependent, they must be large to
have a noticeable effect within the considered energy band. 

Alternatively,  \citet{Boutloukos2010} suggested that, instead of
a blackbody, the spectra of a superburst are better described by a
Bose-Einstein distribution, which is identical to a blackbody with
a deficit at low energies. The observed spectra do not discriminate
between this model and strong local absorption.

\subsection{Comparison to 4U~1820--30's Superburst}

\label{sub:Evidence-for-a}

Much of the observed behavior is qualitatively similar to the 1999
superburst from 4U~1820--30 \citep{Strohmayer2002,Ballantyne2004}.
Both sources are at a similar distance --- $6\,\mathrm{kpc}$ for
4U~1636--536 \citep{Galloway2008catalog} and $6.4\,\mathrm{kpc}$
for 4U~1820--30 \citep{Vacca1986,Ballantyne2005} --- and both superbursts
were observed with $3$ PCUs. 4U~1820--30's superburst, however,
was intrinsically brighter, as it reached the Eddington limit. The
brightness of the reflection signal combined with the helium-rich
disk of 4U~1820--30 produced a stronger Fe K$\alpha$ line, and the
width of the line was used to measure the location on the disk where
reflection originated. This measurement is not possible for 4U~1636--536,
where we have no direct constraints on the location of the reflecting
surface. Conversely, as the blackbody is less bright for 4U~1636--536,
the evolution of the persistent spectral component can be traced during
its superburst \citep{Keek2014sb1}, which was not possible for 4U~1820--30
\citep{Strohmayer2002}. 

Despite the fact that for 4U~1636--536 the peak luminosity is only
$\sim0.2$ of the value for 4U~1820--30, the ionization transition
occurs for both sources on a timescale of $\sim10^{3}\mathrm{s}$:
around $t\simeq6.1\times10^{3}\,\mathrm{s}$ for 4U~1636--536, and
for 4U~1820--30 at $t\simeq1.2\times10^{3}\,\mathrm{s}$. Evacuation
of the inner disk is expected to take place on a dynamical timescale
that is unresolved by our analysis. If present, the effect is smaller
for 4U~1636--536, as its luminosity at the superburst onset is lower.
Processes that take place on a timescale of $\sim10^{3}\,\mathrm{s}$
are likely viscous in nature. For example, puffing-up of the accretion
disk by X-ray heating was a favored mechanism for 4U~1820--30 \citep{Ballantyne2005}.
As no strong correlation of reflection and illumination parameters
are apparent, the timescale of the evolution of the reflection features
appear less dependent on the brightness of the superburst, and more
on innate viscous processes of the accretion disk.

Both superbursts exhibited a similar increase in $N_{\mathrm{H}}$.
4U~1820--30 reached the Eddington limit, and a strong wind from the
neutron star surface or from the irradiated disk could have formed
the absorbing material \citep{Ballantyne2005}. It is interesting
that the weaker superburst from 4U~1636--536 could have produced
an absorber with an $N_{\mathrm{H}}$ of similar magnitude. For both
superbursts $N_{\mathrm{H}}$ remained high in the tail of the superburst,
long after the blackbody flux peaked, indicating a long-lived effect.

\section{Conclusions}

We have shown that the features identified in the spectra of the 2001
superburst from 4U~1636--536 \citep{Keek2014sb1} are consistent
with reflection of the superburst off the photoionized accretion disk.
This is only the second superburst for which this could be detected,
and shows that much of the behavior is also present in a superburst
of typical brightness, as opposed to the particularly powerful superburst
from 4U~1820--30 \citep{Ballantyne2004}. Our spectral fits suggest
that initially reflection occurs off a highly ionized part of the
disk at a larger radius, and transitions in the tail to reflection
off the disk's inner region when its ionization state is lower. The
evolution of the reflecting regions depends on how the intense X-ray
irradiation changes the disk's geometry and ionization profile with
time, and it will require numerical modeling to investigate in detail.
Similar and improved observations of superburst spectra depend on
future missions with large collecting areas such as \emph{NICER }\citep{Gendreau2012NICER}
and \emph{LOFT} \citep{Feroci2014LOFT}. The substantial and evolving
contribution of reflection to the spectrum that we observe for superbursts,
may also be important for the interpretation of short Type I X-ray
bursts. Their spectra are employed to constrain the neutron star equation
of state, but in many cases a changing accretion environment is suspected
to interfere with these measurements \citep[e.g.,][]{Poutanen2014}.
Future instruments can use reflection to trace such changes, and accurately
measure the neutron star mass and radius.

\acknowledgements{LK and DRB acknowledge support from NASA ADAP grant NNX13AI47G and
NSF award AST 1008067. LK \& EK are members of an International Team
on thermonuclear bursts hosted by ISSI in Bern, Switzerland.}

\bibliographystyle{apj}

\begin{thebibliography}{}
\expandafter\ifx\csname natexlab\endcsname\relax\def\natexlab#1{#1}\fi

\bibitem[{{Arnaud}(1996)}]{Arnaud1996}
{Arnaud}, K.~A. 1996, in ASP Conf. Ser. 101: Astronomical Data Analysis
  Software and Systems V, ed. G.~H. {Jacoby} \& J.~{Barnes}, 17

\bibitem[{{Ballantyne}(2004)}]{Ballantyne2004models}
{Ballantyne}, D.~R. 2004, \mnras, 351, 57

\bibitem[{{Ballantyne} \& {Everett}(2005)}]{Ballantyne2005}
{Ballantyne}, D.~R., \& {Everett}, J.~E. 2005, \apj, 626, 364

\bibitem[{{Ballantyne} {et~al.}(2012){Ballantyne}, {Purvis}, {Strausbaugh}, \&
  {Hickox}}]{Ballantyne2012models}
{Ballantyne}, D.~R., {Purvis}, J.~D., {Strausbaugh}, R.~G., \& {Hickox}, R.~C.
  2012, \apjl, 747, L35

\bibitem[{{Ballantyne} \& {Strohmayer}(2004)}]{Ballantyne2004}
{Ballantyne}, D.~R., \& {Strohmayer}, T.~E. 2004, \apjl, 602, L105

\bibitem[{{Blackman}(1999)}]{Blackman1999}
{Blackman}, E.~G. 1999, \mnras, 306, L25

\bibitem[{{Boutloukos} {et~al.}(2010){Boutloukos}, {Miller}, \&
  {Lamb}}]{Boutloukos2010}
{Boutloukos}, S., {Miller}, M.~C., \& {Lamb}, F.~K. 2010, \apjl, 720, L15

\bibitem[{{Bradt} {et~al.}(1993){Bradt}, {Rothschild}, \& {Swank}}]{Bradt1993}
{Bradt}, H.~V., {Rothschild}, R.~E., \& {Swank}, J.~H. 1993, \aaps, 97, 355

\bibitem[{{Casares} {et~al.}(2006){Casares}, {Cornelisse}, {Steeghs},
  {Charles}, {Hynes}, {O'Brien}, \& {Strohmayer}}]{Casares2006}
{Casares}, J., {Cornelisse}, R., {Steeghs}, D., {et~al.} 2006, \mnras, 373,
  1235

\bibitem[{{Cornelisse} {et~al.}(2000){Cornelisse}, {Heise}, {Kuulkers},
  {Verbunt}, \& {in~'t~Zand}}]{Cornelisse2000}
{Cornelisse}, R., {Heise}, J., {Kuulkers}, E., {Verbunt}, F., \& {in~'t~Zand},
  J.~J.~M. 2000, \aap, 357, L21

\bibitem[{{Cumming} \& {Bildsten}(2001)}]{Cumming2001}
{Cumming}, A., \& {Bildsten}, L. 2001, \apjl, 559, L127

\bibitem[{{Fabian} {et~al.}(1989){Fabian}, {Rees}, {Stella}, \&
  {White}}]{Fabian1989}
{Fabian}, A.~C., {Rees}, M.~J., {Stella}, L., \& {White}, N.~E. 1989, \mnras,
  238, 729

\bibitem[{Feroci {et~al.}(2014)Feroci, den Herder, Bozzo, Barret, Brandt,
  Hernanz, \& et~al.}]{Feroci2014LOFT}
Feroci, M., den Herder, J.~W., Bozzo, E., {et~al.} 2014, in Space Telescopes
  and Instrumentation 2014: Ultraviolet to Gamma Ray, ed. {T.~Takahashi,
  J.-W.~A.~den~Herder, \& M.~Bautz}, Vol. 9144, 91442T

\bibitem[{{Fujimoto}(1988)}]{fujimoto88apj}
{Fujimoto}, M.~Y. 1988, \apj, 324, 995

\bibitem[{{Galloway} {et~al.}(2008){Galloway}, {Muno}, {Hartman}, {Psaltis}, \&
  {Chakrabarty}}]{Galloway2008catalog}
{Galloway}, D.~K., {Muno}, M.~P., {Hartman}, J.~M., {Psaltis}, D., \&
  {Chakrabarty}, D. 2008, \apjs, 179, 360

\bibitem[{{Garc{\'{\i}}a} {et~al.}(2013){Garc{\'{\i}}a}, {Dauser}, {Reynolds},
  {Kallman}, {McClintock}, {Wilms}, \& {Eikmann}}]{Garcia2013}
{Garc{\'{\i}}a}, J., {Dauser}, T., {Reynolds}, C.~S., {et~al.} 2013, \apj, 768,
  146

\bibitem[{{Gendreau} {et~al.}(2012){Gendreau}, {Arzoumanian}, \&
  {Okajima}}]{Gendreau2012NICER}
{Gendreau}, K.~C., {Arzoumanian}, Z., \& {Okajima}, T. 2012, in Society of
  Photo-Optical Instrumentation Engineers (SPIE) Conference Series, Vol. 8443

\bibitem[{{Jahoda} {et~al.}(2006){Jahoda}, {Markwardt}, {Radeva}, {Rots},
  {Stark}, {Swank}, {Strohmayer}, \& {Zhang}}]{Jahoda2006}
{Jahoda}, K., {Markwardt}, C.~B., {Radeva}, Y., {et~al.} 2006, \apjs, 163, 401

\bibitem[{{Kallman} \& {Bautista}(2001)}]{Kallman2001xstar}
{Kallman}, T., \& {Bautista}, M. 2001, \apjs, 133, 221

\bibitem[{{Keek} {et~al.}(2014){Keek}, {Ballantyne}, {Kuulkers}, \&
  {Strohmayer}}]{Keek2014sb1}
{Keek}, L., {Ballantyne}, D.~R., {Kuulkers}, E., \& {Strohmayer}, T.~E. 2014,
  \apj, 789, 121

\bibitem[{{Kuulkers} {et~al.}(2004){Kuulkers}, {in~'t~Zand}, {Homan}, {van
  Straaten}, {Altamirano}, \& {van der Klis}}]{2004Kuulkers}
{Kuulkers}, E., {in~'t~Zand}, J., {Homan}, J., {et~al.} 2004, in AIP Conf.
  Proc. 714: X-ray Timing 2003: Rossi and Beyond, 257--260

\bibitem[{{Lapidus} \& {Sunyaev}(1985)}]{lapidus85mnras}
{Lapidus}, I.~I., \& {Sunyaev}, R.~A. 1985, \mnras, 217, 291

\bibitem[{{Lewin} {et~al.}(1993){Lewin}, {van Paradijs}, \& {Taam}}]{Lewin1993}
{Lewin}, W.~H.~G., {van Paradijs}, J., \& {Taam}, R.~E. 1993, Space Science
  Reviews, 62, 223

\bibitem[{{Pandel} {et~al.}(2008){Pandel}, {Kaaret}, \& {Corbel}}]{Pandel2008}
{Pandel}, D., {Kaaret}, P., \& {Corbel}, S. 2008, \apj, 688, 1288

\bibitem[{{Pedersen} {et~al.}(1981){Pedersen}, {van Paradijs}, \&
  {Lewin}}]{Pedersen1981}
{Pedersen}, H., {van Paradijs}, J., \& {Lewin}, W.~H.~G. 1981, \nat, 294, 725

\bibitem[{{Poutanen} {et~al.}(2014){Poutanen}, {N{\"a}ttil{\"a}}, {Kajava},
  {Latvala}, {Galloway}, {Kuulkers}, \& {Suleimanov}}]{Poutanen2014}
{Poutanen}, J., {N{\"a}ttil{\"a}}, J., {Kajava}, J.~J.~E., {et~al.} 2014,
  \mnras, 442, 3777

\bibitem[{{Strohmayer} \& {Bildsten}(2006)}]{Strohmayer2006}
{Strohmayer}, T., \& {Bildsten}, L. 2006, {New views of thermonuclear bursts}
  (Compact stellar X-ray sources), 113--156

\bibitem[{{Strohmayer} \& {Brown}(2002)}]{Strohmayer2002}
{Strohmayer}, T.~E., \& {Brown}, E.~F. 2002, \apj, 566, 1045

\bibitem[{{Strohmayer} \& {Markwardt}(2002)}]{Strohmayer2002a}
{Strohmayer}, T.~E., \& {Markwardt}, C.~B. 2002, \apj, 577, 337

\bibitem[{{Vacca} {et~al.}(1986){Vacca}, {Lewin}, \& {van
  Paradijs}}]{Vacca1986}
{Vacca}, W.~D., {Lewin}, W.~H.~G., \& {van Paradijs}, J. 1986, \mnras, 220, 339

\end{thebibliography}

\end{document}